# The effect of short-range interaction and correlations on the charge and electric field distribution in a model solid electrolyte

T. Patsahan[1], G. Bokun[2], D. di Caprio[3], M. Holovko[1], V. Vikhrenko[2]

[1] Institute for Condensed Matter Physics of the National Academy of Sciences of Ukraine, 1 Svientsitskii Str., 79011 Lviv, Ukraine
[2] Belarusian State Technological University, 13a Sverdlov Str., 220006 Minsk, Belarus
[3] Chimie ParisTech, PSL Research University, CNRS, Institut de Recherche de Chimie Paris (IRCP), 75005 Paris, France

**Abstract**

A simple lattice model of a solid electrolyte presented as a xy-slab geometry system of mobile cations on a background of energetic landscape of the host system and a compensating field of uniformly distributed anions is studied. The system is confined in the z-direction between two oppositely charged walls, which are in parallel to xy-plane. Besides the long-range Coulomb interactions appearing in the system, the short-range attractive potential between cations is considered in our study. We propose the mean field description of this model and extend it by taking into account correlation effects at short distances. Using the free energy minimization at each of z-coordinates, the corresponding set of non-linear equations for the chemical potential is derived. The set of equations was solved numerically with respect to the charge density distribution in order to calculate the cations distribution profile and the electrostatic potential in the system along z-direction under different conditions. An asymmetry of charge distribution profile with respect to the midplane of the system is observed. The effects of the short-range interactions and pair correlations on the charge and electric field distributions are demonstrated.

## 1. Introduction

Solid electrolytes remain an area of intensive scientific activity [1–6] due to their great potential in industrial applications like rechargeable batteries [7,8], fuel cells [9,10], supercapacitors [11], memory devices [12], to name a few. In many cases, solid electrolytes are ceramics or polycrystallites, which require the development of various methods for their production and experimental investigations. The theoretical studies of such materials also encounter serious complications due to an important role played by intergrain boundaries and near-electrode regions.

The most reliable approaches can be based on Monte Carlo or molecular dynamics simulations combined with the quantum density functional theory for the force fields calculations [13–16]. However, these approaches demand high computational resources and suffer from uncertainties arising from the approximations used [17,18]. Additional difficulties are stipulated by inhomogeneities caused by the granular structure of materials and the long-range Coulomb interactions between ions that require to consider systems of very large number of particles. Thus, simplified models of solid electrolytes are frequently used to understand important features of these complicated materials.

Polarization effects at electrodes, perturbation of charge and electric potential distribution near intergrain boundaries strongly affect electro-physical characteristics of devices, where the solid electrolyte is an important constituent element. It was understood for a long time ago that peculiarities of the space charge distribution around these inhomogeneities are determinative for the kinetics of such systems [19–21]. The core-space charge model [22,23] based on the assumptions of dilute defects on a continuum level and without changes of bulk electrolyte characteristics up to the interface gave possibility to explain the charge distribution near the two-phase boundary.

The space charge model (SCM) was applied to investigate grain-boundary phenomena, and it was shown that the enhancement of the grain boundary resistance originates from the charge density depletion in the grain boundary region [20,24,25] that in turn is conditioned by fixed charges in the grain boundary core. Particular applications of SCM are based on specific assumptions about the distribution of the fixed charges in the grain boundary core and in the grains bulk, which have to be justified for materials under investigation [3,6,26–30]. In general, the charge distribution around of space inhomogeneities in ionic systems can result in strong intrinsic electric fields stimulating charge transfer through them [31].

Charge transfer processes in solid electrolytes are frequently described by the transport equations for continuum models [32–34]. On this basis, the linear diffusion model for the grain boundary resistance was suggested and used for a calculation of the current–voltage characteristics of blocking grain boundaries in some oxygen-ion and proton conductors [35–37]. The model reproduces experimentally observed power dependence of current on voltage and results in considerably smaller intergrain energy barrier heights as compared to the results of the conventional SCM.

The concentration of mobile charges in solid electrolytes is usually high enough for energy storage and other applications, therefore the characteristic Debye length is of the order of the lattice spacing in the material under consideration. At these conditions the discrete crystalline structure can affect physico-chemical characteristics and processes in the materials. Lattice models are microscopic in their origin and appropriate for the description of these discrete features [38–46]. They give a possibility to consider systems beyond the assumptions of continuum models and to take into account a wide spectrum of interparticle interactions as well as to predict correlation effects using the statistical mechanics. It was shown that the voltage dependence of the differential capacitance obtained within the mean field approximation for the case of short-range interactions between ions is closer to experimental findings [47,48]. The importance of short-range interactions in a lattice model of ionic systems was demonstrated in [49], where the effect of competition between long-range Coulomb and short-range dispersion interaction on phase transitions was studied. Recently, the short-range dispersion (or Van der Waals) interactions were addressed to explain some experimentally observed peculiarities of electrical double layer capacitance in ionic liquid-solvent mixtures [50]. With respect to solid electrolytes, lattice models provide the ability to describe their structural inhomogeneities [51].

In this paper, we investigate the effect of short-range attractive interaction between mobile ions on charge and electric potential distribution near charged planes or intergrain boundaries modeled by variations of the host system energy landscape. The short-range interactions can originate from potentials of different nature, e.g. Van der Waals attraction or strain induced repulsion. Electrostatic interactions are repulsive between ions of the same sign. Thus attractive short-range interactions being opposite to electrostatic ones can have a stronger effect on the electrical characteristics of the system. Alongside with this interparticle correlations are taken into account through approximate correlation functions. The theory is represented in the next section. The calculation results and their discussion are presented in the third section. The conclusions are drawn in the last section.

## 2. The constitutive equations for concentration and electric field distribution

We study a simple-cubic lattice model of a solid electrolyte presented as a xy-slab geometry system of mobile cations on a background of energetic landscape of the host system and in a compensating field of uniformly distributed anions [52], which are fixed at their positions [51]. Therefore, the inhomogeneities appearing in the system concern only cations species. The anions are taken into account implicitly by creating a constant background of some charge density, which neutralizes the total charge in the system. We also consider a presence of two oppositely charged walls parallel to xy-plane confining the system in the z-direction. Consequently, the distribution of cations is affected simultaneously by the energetic landscape and the external electric field produced by the charged walls. Charged impurities segregated in the grain boundary core can easily be taken into account in our approach; however, we concentrate our attention on energetic inhomogeneities in the intergrain region that can significantly influence the charge and electric field distribution in the system. Besides long-range electrostatic interactions, the short-range cation-cation interactions is introduced in our model.

To describe the model presented in our study we propose the approach based on the mean field approximation (MF) and its extension, which takes into account the short-range cation-cation correlations (MF+corr). We are aimed to calculate the charge distribution profile $\rho(z_i)$ and the electrostatic potential $\psi(z_i)$ in the system along z-direction under different conditions. For this purpose, using the free energy minimization at each of z-coordinates we have derived the corresponding set of non-linear equations $\mu_i[\rho(z_i)] = \mu_r$, which are solved numerically with respect to the particle density $\rho(z_i)$. Having the charge distribution in the system we obtain the profile of electric field $E(z_i)$ and the electrostatic potential $\psi(z_i)$.

The minimization of the free energy of the system that is equivalent to the approach of Ref. [51] leads to the following equations for the chemical potential of positively charged particles (cations subsystem)

$$\beta\mu_i - \beta\mu_r = 0, \qquad (1)$$

where the index $i$ is considered in the range $i = 1,...,n_z$, the total chemical potential $\beta\mu_i$ at $z_i = a \cdot i$ is defined as

$$\beta\mu_i = \beta\mu_i^{id} + \beta\mu_i^{sr} + \beta\mu_i^{el} + \beta\mu_i^{f} + \beta\mu_i^{g} \qquad (2)$$

and $\beta\mu_r$ is the chemical potential of the reservoir, $\beta = 1/k_B T$ is the inverse temperature, $k_B$ is the Boltzmann constant, $T$ is the temperature, $a$ is the lattice spacing.

The ideal gas term in equation (2) is

$$\beta\mu_i^{id} = \ln\frac{c_i}{1-c_i}, \qquad (3)$$

where $c_i = \rho(z_i)a^3$ is the reduced number density of cations.

The terms $\beta\mu_i^{sr}$ and $\beta\mu_i^{el}$ appearing due to the short-range and electrostatic interactions between cations are considered as the part of internal energy, which can be calculated in accordance with the standard relation $\beta\mu_i^{ex} = \beta\int \rho(\mathbf{r}_j)u(r_{ij})g(r_{ij})d\mathbf{r}_j$, where $u(r_{ij})$ is the pair potential of interaction between particles and $g(r) = 1 + h(r)$ is the pair distribution function [53-55]. Since the dispersion short-range interaction decays rapidly, we limit this type of interaction by the first coordination sphere. In the case of simple-cubic lattice model the short-range interaction between particles in the first coordination sphere contains 6 sites and leads to the following expression for $\beta\mu_i^{sr}$:

$$\beta\mu_i^{sr} = \beta\mu_i^{sr,mf} + \beta\mu_i^{sr,cor} = \beta J c_{i-1}(1 + h_{i,i-1}^{(1)}) + 4\beta J c_i(1 + h_{i,i}^{(1)}) + \beta J c_{i+1}(1 + h_{i,i+1}^{(1)}), \qquad (4)$$

where $\beta\mu_i^{sr,mf} = \beta J c_{i-1} + 4\beta J c_i + \beta J c_{i+1}$ is the mean-field (MF) contribution (that gives birth to '1' in the parentheses of the rhs of Eq. (4)) and $\beta\mu_i^{sr,cor}$ takes into account the pair correlations $h_{i,j}^{(1)}$ between cations on nearest neighbor sites at the distance of lattice spacing $a$. The lower indices $i, j$ denote lattice plains perpendicular to $z$ axis. It should be noted, that the short-range interactions are summed over the four equivalent nearest neighbor sites in the same plain and two sites in the nearest plains. The negative value of the interaction constant $J$ corresponds to the attractive interparticle interaction.

The electrostatic term we calculate from the solution of the Poisson equation written for the model considered in our study as

$$\nabla^2 \psi(z_i) = -\frac{q}{\varepsilon_0 \varepsilon}(c(z_i) - c), \qquad (5)$$

$$c = \frac{1}{n_z}\sum_{k=1}^{n_z} c_k, \qquad (6)$$

where $c$ is the average concentration of cations in the system, which is equivalent to the concentration of negatively charged background $c_-$ according to the electro-neutrality condition. We would like to recall that the negative charge is distributed homogeneously in the system, hence the $c_- = c$ at any $z_i$.

The solution of Eq. (5) with respect to the electric potential $\psi(z_i)$ gives the mean-field contribution to the chemical potential $\beta\mu_i^{el}$, which can be presented in the following form

$$\beta\mu_i^{el,mf} = \beta\psi_i q = -\frac{2\pi\lambda_B}{a}\sum_{j=1}^{n_z}\langle c_j - c\rangle|j - i| \qquad (7)$$

where $\lambda_B = q^2/(4\pi\varepsilon_0\varepsilon k_B T)$ is the Bjerrum length, $q$ is the elementary charge, $\varepsilon_0$ is the vacuum permittivity, $\varepsilon$ is the dielectric constant assumed to be constant throughout the entire system. The distance between the walls is defined as $L_z = (n_z + 1)a$.

As for the short-range term, we take into account the pair correlation effects induced by the nearest neighbors due to the electrostatic interaction. But in this case, except the first coordination sphere, we also consider the second coordination sphere located at the distance $a\sqrt{2}$ (contains 12 sites) and the third coordination sphere located at the distance $a\sqrt{3}$ (contains 8 sites). Therefore, the term of chemical potential due to the electrostatic interaction, $\beta\mu_i^{el}$, together with the correlation contribution, $\beta\mu_i^{el,corr}$, can be expressed as

$$\beta\mu_i^{el} = \beta\mu_i^{el,mf} + \beta\mu_i^{el,corr} = -\frac{2\pi\lambda_B}{a}\sum_{j=1}^{n_z}\langle c_j - c\rangle |j-i| - \frac{2\pi\lambda_B}{a}\left(c_{i-1}h_{i,i-1}^{(1)} + 4c_ih_{i,i}^{(1)} + c_{i+1}h_{i,i+1}^{(1)}\right)$$
$$-\frac{2\pi\lambda_B}{a\sqrt{2}}\left(4c_{i-1}h_{i,i-1}^{(2)} + 4c_ih_{i,i}^{(2)} + 4c_{i+1}h_{i,i+1}^{(2)}\right) - \frac{2\pi\lambda_B}{a\sqrt{3}}\left(4c_{i-1}h_{i,i-1}^{(3)} + 4c_{i+1}h_{i,i+1}^{(3)}\right), \quad (8)$$

where the upper index in the pair correlation functions $h_{ij}^{(k)}$ denotes the coordination sphere (i.e. $k=1$, 2 or 3) and the lower index denotes the lattice plain where the site is disposed.

We consider the cation-cation pair correlation functions $h_{ij}^{(k)}$ in the simplified way using the definition of the potential of mean force (PMF) [53-55], which in the lattice representation looks as

$$h_{ij}^{(k)} = g_{ij}^{(k)} - 1 = \exp\left(-\beta w_{ij}^{(k)}\right) - 1, \quad (9)$$

The PMF $w_{ij}^{(k)}$ consists of the short-range attraction $J$ and the screened Coulomb potential with the Debye length $\lambda_D = 1/\kappa$ [53-55]

$$\beta w_{ij}^{(1)} = \beta J + \frac{\lambda_B}{a}e^{-\kappa a}, \quad \beta w_{ij}^{(2)} = \frac{\lambda_B}{a\sqrt{2}}e^{-\kappa a\sqrt{2}}, \quad \beta w_{ij}^{(3)} = \frac{\lambda_B}{a\sqrt{3}}e^{-\kappa a\sqrt{3}}. \quad (10)$$

According to the model the short-range interaction is cut off at the first six nearest neighbor sites belonging to the first coordination sphere, hence the term $J$ related to this type of interaction is present only in $w_{ij}^{(1)}$. The inverse Debye screening length in (10) is expressed according to [41,51] as $\kappa = \sqrt{4\pi\lambda_B c(1-c)/a^3}$. For the temperature and concentrations considered in this study, the Debye screening length is of the order of $a$. In this case, the screened long-range Coulomb interaction has to be taken into account up to the third coordination sphere, thus included in all considered $w_{ij}^{(k)}$.

The next term $\beta\mu_i^f$ in the chemical potential appears when the external electric field is applied. We consider this field as the one induced by the oppositely charged walls with the charge density $qc_w/a^2$. This term can be derived by calculating the interaction energy of a probe cation particle on the lattice plain $i$ with both charged walls:

$$\beta\mu_i^f = -\frac{4\pi\lambda_B c_w}{a}\left(i - \frac{n_z+1}{2}\right), \quad (11)$$

where $c_w$ is the dimensionless charge density at each of the walls.

Finally, the last term in (2) describes the presence of intergrain boundaries, which are modelled by the deviation of the energetic landscape from the bulk value as

$$\beta\mu_i^g = J_g, \text{ when } i = \frac{n_z}{2} \text{ or } i = \frac{n_z}{2} + 1. \quad (12)$$

A positive value of the energy deviation $J_g$ corresponds to a lower probability for cations to occupy the lattice sites belonging to the grain boundaries.

Therefore, taking into account all of the terms of the chemical potential $\mu_i$ mentioned above, the set of equations (1) is solved numerically with respect to $c_i$ for $i=1,..,n_z$. For this purpose, the Newton-Raphson algorithm have been applied in a combination with an additional iterative procedure needful to adjust the free parameter $\mu_r$ in such a way to get the average cation concentration equal to the certain values of $c$ used in our study. The accuracy of such an adjustment have been restricted by $10^{-9}$. Due to the confinement the system boundary conditions are set as $c_0 = 0$ and $c_{n_z+1} = 0$.

## 3. Results and discussion

The numerical results can be addressed to the system of real charged particles like Li in Li-ion conductors or crystalline defects exploring the Kröger–Vink notation [40] (e.g., for high temperature solid electrolytes). However, the aim of the paper is to study the general features of effects caused by the electric field and short-range interionic interaction on the charge and electric field distribution, and how they behave in the different approximations. The calculations were done for two ionic or defect average concentrations of cations $c = 0.03$ and $0.1$ for the system with the dielectric constant $\varepsilon = 41.8$ and the lattice spacing $a = 0.4$ nm, with the Bjerrum length $\lambda_B = a$ that corresponds to $T = 1000$ K. The Debye length is equal to $0.94a$ and $1.65a$ for the lower and higher concentrations, respectively. Some of quantities used in our paper are presented in reduced units, such as the reduced distance to the left wall is $z_i^* = z_i / a$ and the concentrations are normalized as $c = \rho a^3$. It is also worth noting that we do not consider the short-range interaction with the walls in the current study, however this interaction was partially analyzed in the previous paper [51].

First, we consider the effect of external electric field on the charge distribution in the system without the intergrain boundary and short-range interaction. In Fig. 1 we present the charge distribution depending on electric field at the lower average concentration $c = 0.03$ and at the different charge densities of the confining walls (electrodes) ranging in $c_w = 0.0 - 0.3$. One can observe that at the left positively charged wall the charge depletion layer grows when the wall charge increases. On the other hand, the cations are accumulated at the opposite wall. The depletion layer appears because of the electroneutrality condition. The lower is the cation concentration the larger volume of the system is required to fulfill this condition.

The effect of correlations is minor at the left wall due to the low cation concentrations and it is more pronounced at the right wall. The correlations lead to nonmonotonous charge distribution that is not observed in the mean field approximation. Moreover, the charge correlations lead to the deviation of the charge density from the average value near the walls even when the walls are not charged. These features become more pronounced at the larger average density $c = 0.10$, when the width of the near-wall structures decreases because of easier fulfilling the electroneutrality condition and due to the smaller Debye length (Fig. 2). The charge oscillations are consequences of the interparticle correlations and appear at not very high charge concentrations far from saturation in contrast to density/charge oscillations in dense liquid electrolytes and ionic liquids [44,56–58]. It is noted in the recent paper [59] that in the mean field approximation such oscillations do not appear for effective attractive short-range interactions.

For the system of cations with the short-range attraction one can observe qualitatively different concentration profiles depending on the approximation applied. It is seen in Fig. 3a that for the case of average concentration $c = 0.03$ and uncharged walls the contact value of cations profile at the both walls is negative within the mean field approximation, while taking into account correlations results in the positive contact value. The same is observed for the larger average concentration $c = 0.10$. However, due to correlation effects the oscillating profile with periodic structure appears along $z$-direction (Fig. 3b). Similar it was found in systems with competing interactions [60,61]. This result is expected, since in our model in fact the competing interaction between cations is introduced through the short-range attraction and the electrostatic repulsion on longer distances.

The behavior of the charge distribution near the grain boundary does not affect the phenomena near the walls and *vice versa* if the system size is considerably larger than the Debye length. At the lower concentration for the system size of 30 layers this condition is not completely fulfilled especially at the large concentrations of wall charge where the depletion zone is sufficiently wide. The charge distribution near the intergrain boundary is slightly asymmetric, and without short-range interactions the correlations only slightly change the results within the mean field approximation (Fig. 4). At the higher concentration, the charge distribution near the grain boundary are not affected by the walls and the correlations reduce the deviation of the charge density from its average value (Fig. 5).

The effect of the short-range attraction is demonstrated in Fig. 6. The interaction within the mean field approximation slightly redistribute the charge density profile near the negatively charged wall and almost does not influence the distribution near the positively charged wall in view of the low mobile particles concentration in this region (Fig. 6a). The charge distribution near the grain boundary does not depend on the walls, while the short-range attraction enhances the deviation of the charge distribution from the average concentration (Fig 6b).

The electric potential profiles throughout the system are shown in Fig. 7a at the different wall charges ($c_w = 0.0 - 0.3$) and in the presence of intergrain boundary. It is seen that in the case of charged walls the electric potential changes almost linearly with *z*. However, small deviations from the linear dependence are noticed near the walls. On the scale of the total potential difference, the perturbation in the region of the grain boundary can hardly be seen. On the other hand, in the case of uncharged walls ($c_w = 0.0$) the potential variations in this region are quite significant (Fig. 7b). When the correlations are taken into account, the pronounce deviation of electric potential is observed both in the intergrain boundary and in the near-wall regions.

The electric field induced by the mobile cations is presented in Fig. 8. One can see that the obtained profiles of electric field along the system are more informative. Besides essential field variations near the walls the effect of the intergrain boundary is visible on the curves. At the distances far enough from the confining walls and from the intergrain boundary the electric field should be equal to the external field, but with the opposite sign. As it was mentioned above, the charge distribution, and accordingly the variations of electric field distribution in the vicinity of grain boundary are not affected by the external field if the system is large enough. Strong perturbations of the electric field in the region of the integrain boundary are seen in Fig. 8b. For the parameters introduced in the beginning of the section the dimensionless electric field $\beta qaE = 0.1$ corresponds to $2.15 \times 10^7$ V/m. Although in equilibrium, the external electric field does not influence the internal electric field in the intergrain boundary, the electric current can be significantly affected by the intrinsic field variations [51].

The qualitative features of the electric field distribution remain similar at the lower concentration (Fig. 9), however the amplitudes of the variations decrease. Taking into account the correlations reduces the electric field variations in the intergrain boundary region and considerably enhances the double layer structure near the charged walls in the both cases.

The situation in real ceramic solid electrolytes is much more complicated as compared to the model used in this study. The main complications arise due to non-uniform distribution of ions of the host subsystem because of their segregation at grain boundaries. These ions are considered as fixed at their positions. This segregation is well documented by experimental findings as well as computer simulation results and explained by binding energetics of grain boundaries [25,27,62–67]. The same binding arguments can be applied to mobile species like oxygen vacancies in oxides or Li ions [66,68,69]. The typical binding energies are of the order of several tenths of eV. In some cases, oxygen ions tend to segregate at the grain boundary while depletion of vacancies is observed at the grain boundary core [69]. This specific situation is considered in the current study with the characteristic energy around of 0.1 eV. The suggested model is quite flexible and can easily incorporate binding of ions by short range Van der Waals type and long range Coulomb interactions and more complicated interactions with the host subsystem of the type of Schottky barriers [27,70,71].

4. Conclusions

A simple model of ceramic solid electrolyte confined between two oppositely charged walls is studied. The intergrain boundary is modelled by the corresponding energy landscape of the host system. The mobile cations are considered on the neutralizing background of fixed anions. The mean field approach for the description of density profiles of cations between oppositely charged walls is proposed. This approach is extended by taking into account correlations between cations at short distances.

It is shown that at strong external electric fields (high charge density of the confining walls) a wide depleting zone near the positively charged wall is observed, which gets narrower if the average cation concentration is increased. Besides the Coulomb interaction the effect of short-range attractive potential is introduced. It was found that the cations accumulate near the negatively charged wall, where the short-range interaction leads to a nonmonotonous charge distribution; however, this behavior can be predicted only with taking into account correlations.

The distribution of charge near the walls and intergrain boundary are independent of each other unless the system size is comparable with the width of the depletion zone and the Debye length. Due to positive values of the site energy deviation at the intergrain boundary with respect to the bulk value an essential deficit of charge carriers appears inside this region that results in the excess charge carriers beyond the intergrain boundary. This is a consequence of the total charge neutrality in the system leading to the complicated electric field distribution in the grain boundary region.

The redistribution of the mobile charges in the grain boundary region due to the variations of energetic landscape of the host material results in complicated profiles of the electric potential and field that should significantly affect the current transfer processes in this system. The short-range attractive interactions and correlations considerably change the numerical values of the parameters characterizing the electric field and potential distribution. The results obtained show that alongside with Coulomb interactions the binding energetics as well as landscape peculiarities can play a crucial role in the charge and electric potential distribution.

**Acknowledgement**

This project has received funding from the European Union's Horizon 2020 research and innovation programme under the Marie Skłodowska-Curie grant agreement No. 734276.

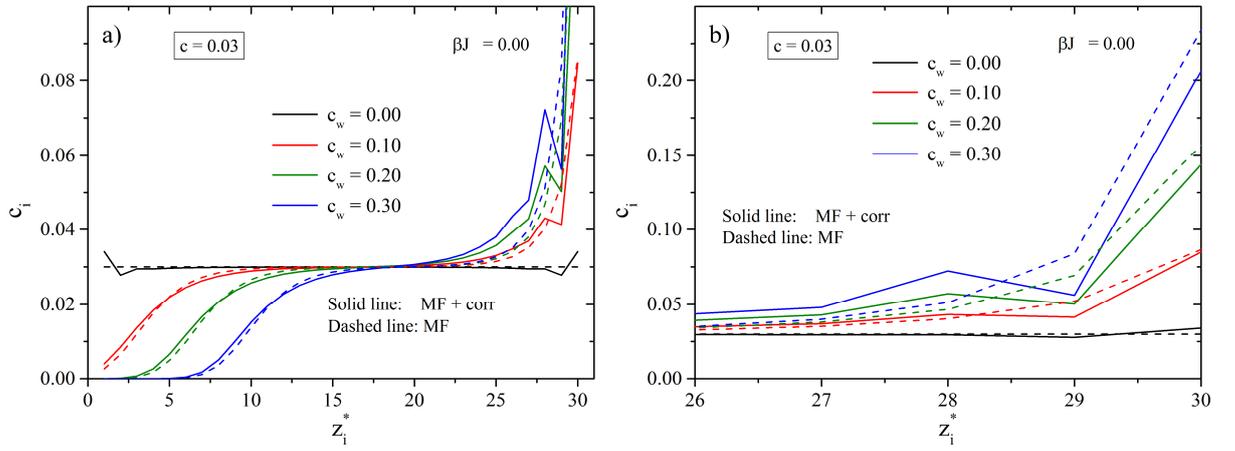

**Fig. 1.** The concentration profile of cations between the walls of different charges ($c_w$=0.0–0.3) at the lower average concentration ($c$=0.03). Dashed lines correspond to the results obtained within the mean field (MF) approximation and solid lines obtained with taking into account correlations (MF+corr). On the right panel, the region near the right wall is shown in a larger scale.

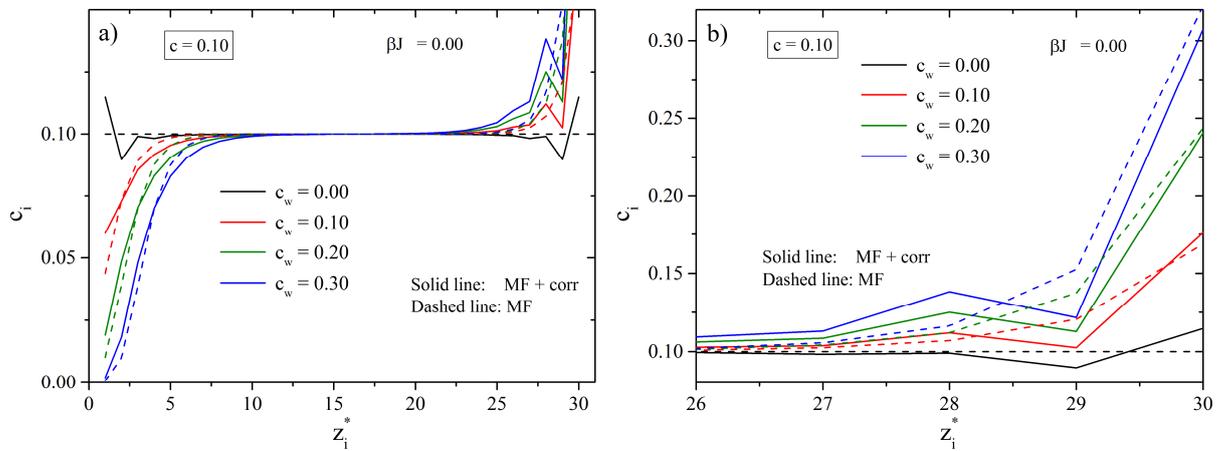

**Fig.2.** The same as in Fig. 1, but at the higher average concentration of cations ($c$=0.10).

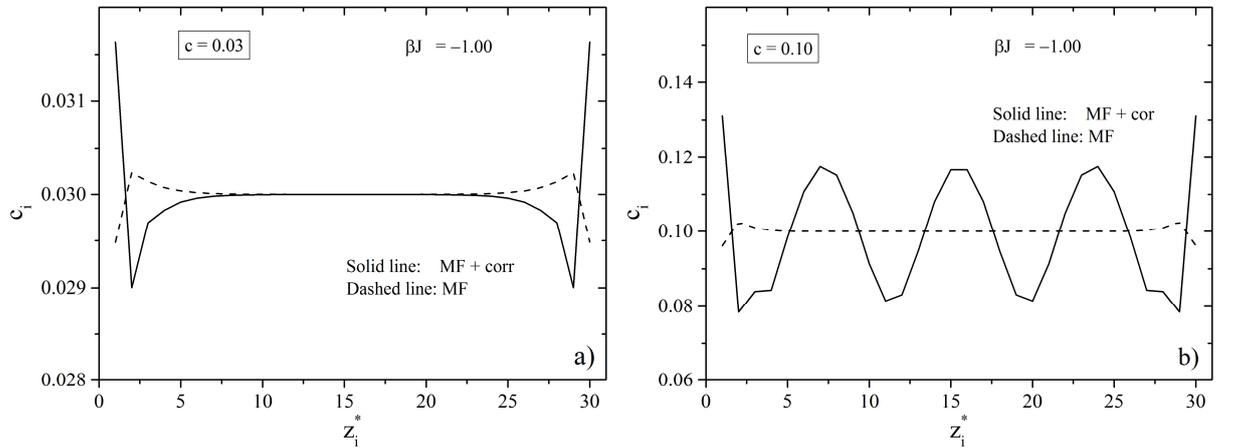

**Fig. 3.** The concentration profile of cations between uncharged walls ($c_w$=0.0), but with taking into account the short-range attraction ($\beta J$=−1.0) at the low (left panel, $c$=0.03) and high (right panel, $c$=0.10) average concentrations. Comparison between the MF (dashed line) and MF+corr (solid line) approximations.

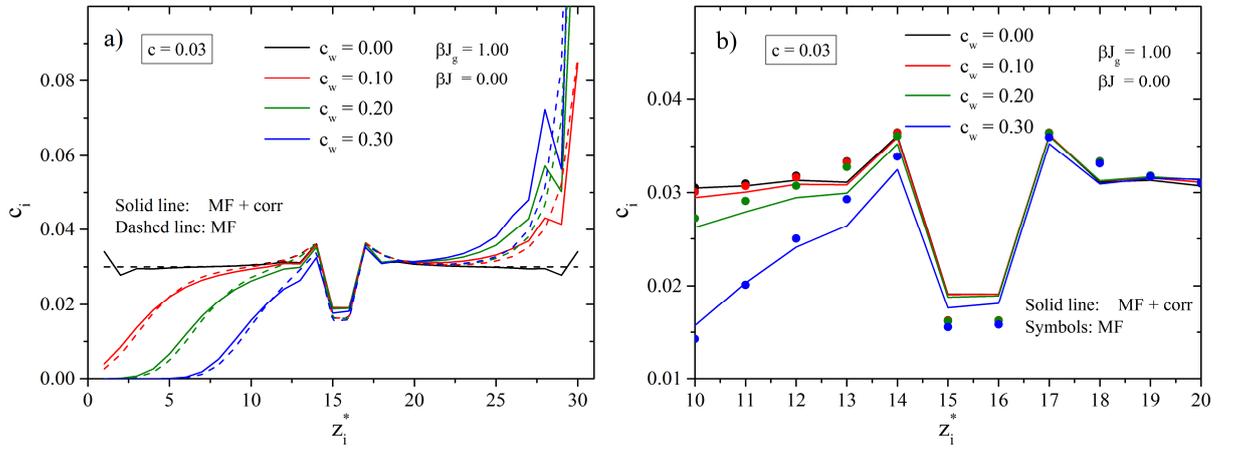

**Fig. 4.** The concentration profile of cations between the walls of different charges ($c_w$=0.0–0.3) at the lower average concentration ($c$=0.03) and with taking into account the intergrain boundaries ($\beta J_g$=1.00), but with no short-range attraction ($\beta J$=0.0). On the right panel, the intergrain region is shown in a larger scale.

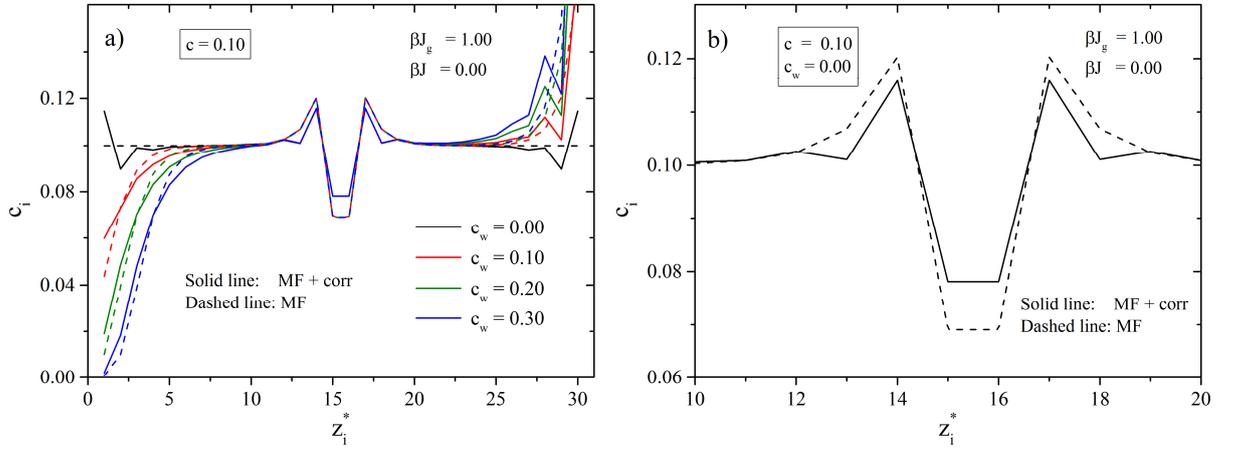

**Fig. 5.** The same as in Fig. 4, but at the high average concentration of cations ($c$=0.10).

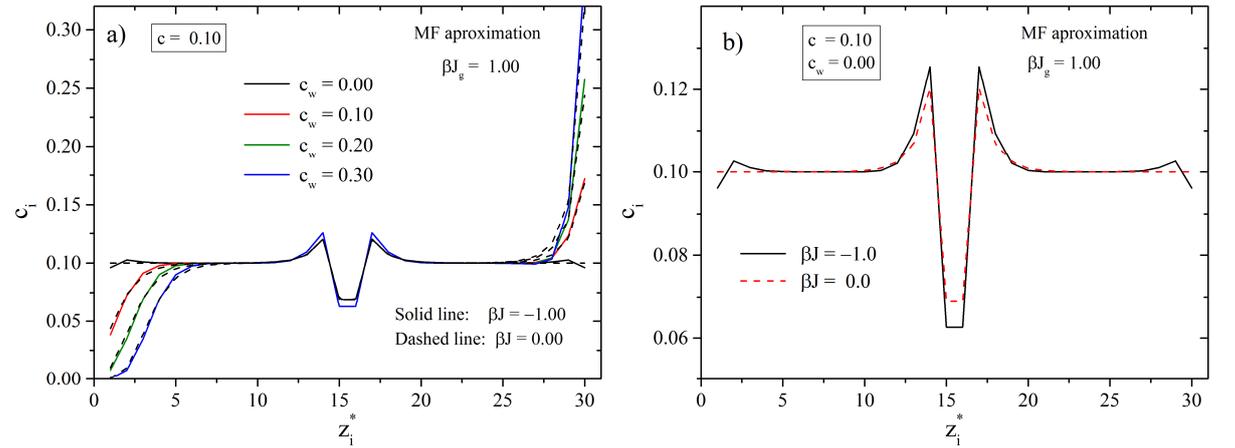

**Fig. 6.** The concentration profile of cations between the wall of different charges ($c_w$=0.0–0.3), but with taking into account the short-range attraction ($\beta J$=–1.0, solid lines) and the intergrain boundaries presence ($\beta J_g$=1.00) at the average concentration $c$=0.10. For a comparison, the results for cations without the short-range attraction ($\beta J$=0.0) are denoted by dashed lines.

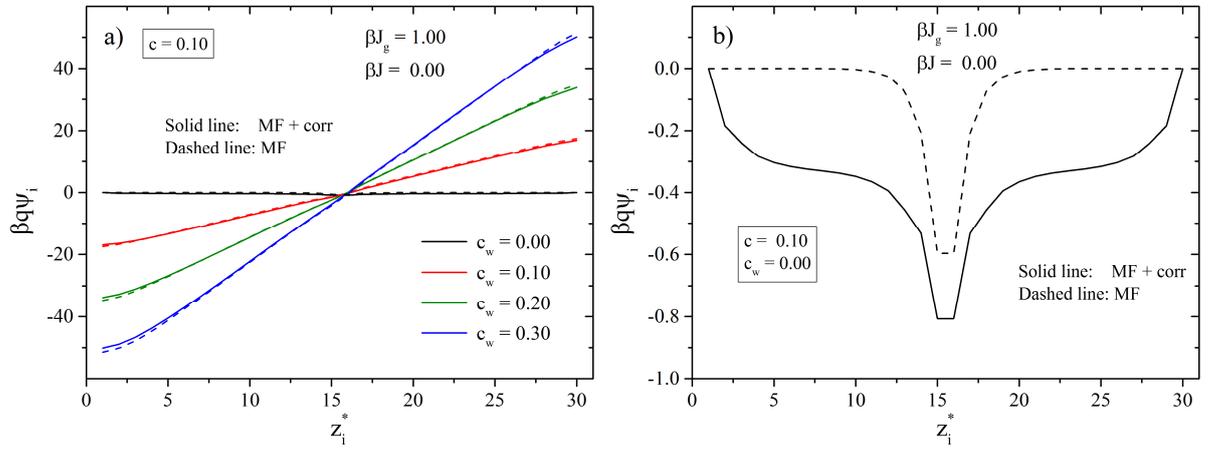

**Fig. 7.** The electric potential profile of cations between the walls of different charges ($c_w = 0.0-0.3$) at the average concentration of cations $c=0.10$ and with taking into account the intergrain boundaries ($\beta J_g=1.00$), but without short-range attraction ($\beta J=0.00$). On the right panel, the intergrain region is shown in a larger scale.

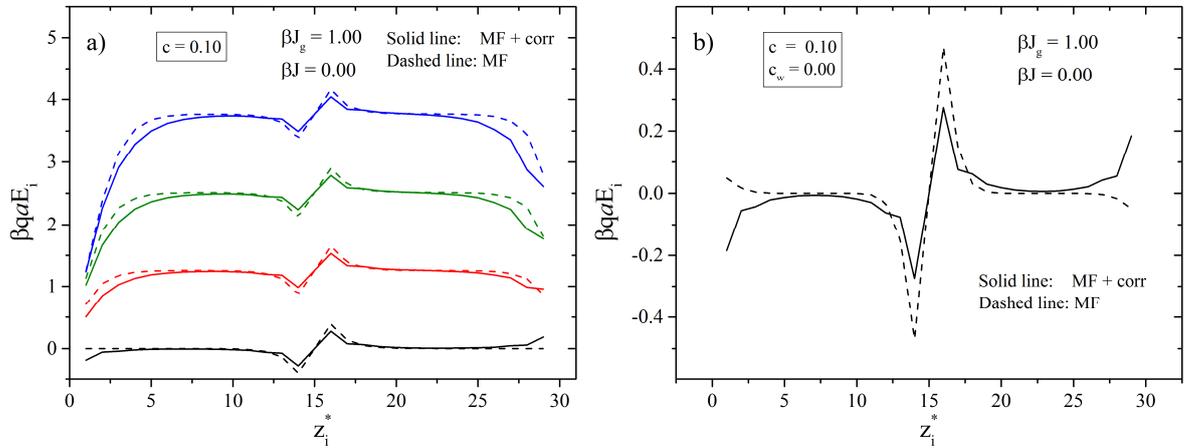

**Fig. 8.** The electric field profile of cations between the walls of different charges ($c_w=0.0-0.3$) at the higher average concentration of cations $c=0.10$ and with taking into account the intergrain boundaries ($\beta J_g=1.00$), but without short-range attraction ($\beta J=0.0$). The colors denote the same values of $c_w$ as in the previous figures. On the right panel, the intergrain region is shown in a larger scale.

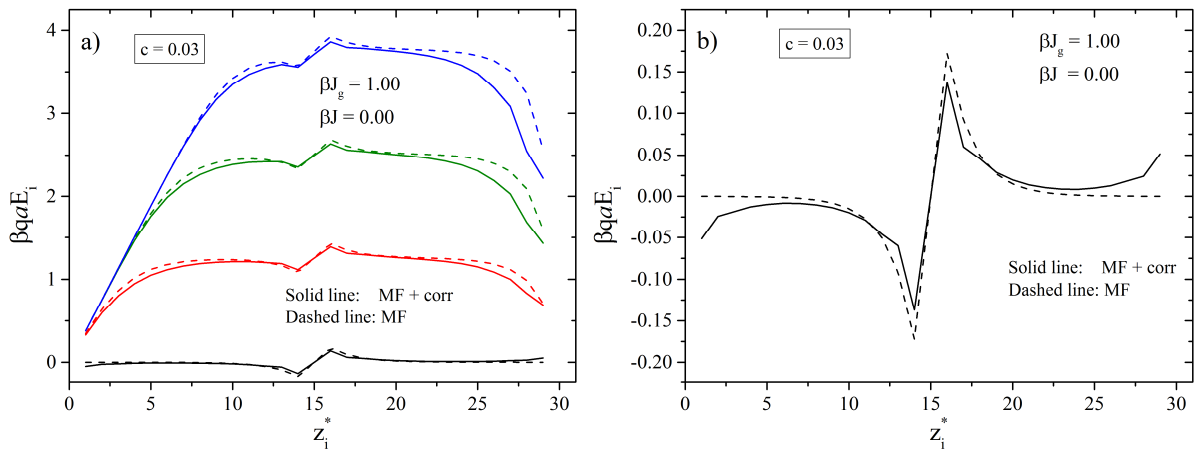

**Fig. 9:** The same as in Fig. 8, but for the lower average concentration of cations ($c=0.03$).